\documentclass[twocolumn,showpacs,preprintnumbers,amsmath,amssymb]{revtex4}
\usepackage{graphicx}
\usepackage{dcolumn}
\usepackage{bm}

\begin{document}
\bibliographystyle{aip}

\title{Defects in Graphene-Based Twisted Nanoribbons:\\ Structural, Electronic and Optical Properties}

\author{E. W. S. Caetano}
\affiliation{Centro Federal de Educa\c{c}\~ao Tecnol\'ogica do Cear\'a, Avenida 13 de Maio, 2081, Benfica, 60040-531 Fortaleza, Cear\'a, Brazil}

\author{V. N. Freire, S. G. dos Santos}
\affiliation{Departamento de F\'{\i}sica, Universidade Federal do Cear\'a, Centro de Ci\^encias, Caixa Postal 6030, Campus do Pici, 60455-760, Fortaleza, Cear\'a, Brazil}

\author{E. L. Albuquerque}
\affiliation{Departamento de F\'{\i}sica Te\'orica e Experimental, Universidade Federal do Rio Grande do Norte, 59072-970 Natal, Rio Grande do Norte, Brazil}

\author{D. S. Galv\~{a}o, F. Sato}
\affiliation{Instituto de F\'{\i}sica Gleb Wataghin, Universidade Estadual de Campinas, Unicamp, 13083-970, Campinas, S\~{a}o Paulo, Brazil}

\date{\today}

\begin{abstract}
We present some computational simulations of graphene-based nanoribbons with a number of half-twists varying from 0 to 4 and two types of defects obtained by removing a single carbon atom from two different sites. Optimized geometries are found by using a mix of classical-quantum semiempirical computations. According with the simulations results, the local curvature of the nanoribbons increases at the defect sites, specially for a higher number of half-twists. The HOMO-LUMO energy gap of the nanostructures has significant variation when the number of half-twists increases for the defective nanoribbons. At the quantum semiempirical level, the first optically active transitions and oscillator strengths are calculated using the full configuration interaction (CI) framework, and the optical absorption in the UV/Visible range (electronic transitions) and in the infrared (vibrational transitions) are achieved. Distinct nanoribbons show unique spectral signatures in the UV/Visible range, with the first absorption peaks in wavelengths ranging from the orange to the violet. Strong absorption is observed in the ultraviolet region, although differences in their infrared spectra are hardly discernible.
\end{abstract}

\maketitle

\section{Introduction}

Most recent advances in nanoscience and nanotechnology were made through the fabrication and characterization of novel carbon-based nanostructures such as the buckminsterfullerene \cite{bf1}, carbon nanotubes \cite{nt1,nt2,nt3} and graphenes \cite{gf1,gf2,gf3}. Recently it is an indisputable fact that carbon has unique chemical properties among all elements of the periodic table, being essential to construct those nanomachines that existed in the natural world before man appearance on Earth and work inside the cells of all living organisms \cite{nm1}.

Fullerenes and carbon nanotubes can be formed by rolling up a graphene sheet around a sphere and a cylinder, respectively. Graphene shows many interesting features, such as extraordinary electronic transport properties \cite{gf4,gf5,gf6,gf7} related to its Dirac-like band structure. Indeed, a graphene sensor was able to detect the adsorption of a single gas molecule thanks to the high sensitivity of its electrical resistance to local changes in carrier concentration \cite{gf8}. Another structure that can be made from a single sheet of graphene is a closed ribbon (one can think of a single wall carbon nanotube as a very wide closed carbon ribbon with a small radius). It is possible to conceive a closed carbon nanoribbon with one or more half-twists along its length. In particular, the nanoribbon with a single half-twist has the same features of the famous M\"{o}bius strip, after the German mathematician August Ferdinand M\"{o}bius (1790-1868) who discovered it. A M\"{o}bius strip is a nonorientable surface in the Euclidean space $\Re^3$, which means that a two-dimensional object transported around the surface can return to the point where it started looking like its image reflected in a mirror. General equilibrium equations for physical twisted strips were recently obtained by Starostin and Van Der Heijden \cite{mob0}.

Molecules with twists and M\"{o}bius topology have already been investigated in the literature, from both the theoretical and experimental viewpoints. More than forty years ago, Heilbronner \cite{mob1} defined the concept of a M\"{o}bius aromaticity for cyclic molecules, predicting the stability of M\"{o}bius aromatic hydrocarbons with 4n p-electrons. Semiempirical and first principles calculations were used to investigate M\"{o}bius annulenes and twisted cyclacenes \cite{mob2,mob3}. In particular, the calculations for cyclacenes with one, two and three half-twists \cite{mob3} revealed a localisation of the twist over 2-4 benzene rings. First principles calculations using the Density Functional Theory framework predicted the existence of a annulene with the topology of a M\"{o}bius strip with two half-twists \cite{mob4}.

In 2003, Ajami \textit{et al.} presented a paper reporting the synthesis of a neutral M\"{o}bius aromatic hydrocarbon \cite{mob5} by combining two systems. The first one consisted in a flat aromatic ring with p-orbitals  perpendicular to its plane, while the second system was a curved aromatic ring with pyramidalized bonds and p-orbitals parallel to the plane of the ring. Both systems were fused chemically through pericyclic reactions and five isomers were obtained, one of them with M\"{o}bius topology. Nevertheless, the several methodological errors pointed in the work of Ajami \textit{et al.} with respect to the assignment of M\"{o}bius aromaticity to their newfound molecule \cite{mob6} -- a critique that deserved a reply \cite{mob7} -- no one contests that a M\"{o}bius topology for a molecule was achieved in it. Other examples of M\"{o}bius structures obtained experimentally are the crystals of NbSe$_3$ synthesized by Tanda \textit{et al.} \cite{mob8} and a protein with M\"{o}bius topology and insecticidal activity found in some plants \cite{mob9}. A recent account on the design of molecules with M\"{o}bius features is presented in Ref.\ \cite{mob10}.

In a previous paper \cite{mob11}, we presented results of classical force field, semiempirical and first principles simulations for twisted graphene nanoribbons with up to 7 half-twists, investigating their structural stabilities, electronic structure and optical properties. In this work we go further by studying the structural features of twisted graphene nanoribbons with defects created by the removal of a single carbon atom from different sites along the ribbon width. This kind of defect was chosen by us due to its size and aspect ratio as the removal of a single atom (single vacancy) is the most expected defect to be present in the nanoribbons. Other defects such as Stone-Wales, which are precursors of fractures in carbon nanotubes \cite{mob11sw}, could also occur but they should be less probable and of higher formation energy. Among the methods for studying the effect of defects in carbon nanostructures one can cite the theory proposed by Wu \textit{et al.} \cite{mob11at} based on the interatomic potential, incorporating the effect of curvature and bending moment for curved surfaces, which allows for the determination of constitutive relations involving the stress, moment, strain and curvature as functions of the interatomic potential. Computational simulations were carried out using classical dynamics and the semiempirical formalism. We show how the presence of defects affects the curvature and other structural characteristics of the nanoribbons, their electronic states and their infrared and UV-VIS spectra. We also perform a comparative analysis of these physical properties as functions of the number of half-twists in the graphene nanoribbons, searching for molecular signatures that could be helpful to experimental testing.

\section{Computational Methodology}

All twisted nanoribbons studied here were based in a single rectangular strip of graphene with armchair and zigzag edges, as in Ref.\ \cite{mob11}. The strip dimensions, $L$ and $W$, are obtained by counting the number of C--C dimers parallel and perpendicular, respectively, to the armchair and zigzag sides. For the present investigation we have chosen $L=39$ and $W=7$, as shown in Fig.\ 1(a), which correspond to a strip length of 8.26 nm and width of 0.727 nm. This strip is larger than the one used in our previous work, $L=29$, $W=5$ \cite{mob11} to allow for the evaluation of two different defect sites, and it is consistent in size and aspect ratio with some recent experimental reports \cite{exp1,exp2,exp3}. The nanostructure has 280 carbon atoms and 86 hydrogen atoms are added to passivate their dangling bonds -- 6 hydrogen atoms at the zigzag edges are removed when the strip ends are joined -- so the chemical formulae of the flat rectangular strip and the closed nanoribbons are C$_{280}$H$_{86}$ and C$_{280}$H$_{80}$. A single defect can be obtained by removing one carbon atom from the strip. We have chosen to perform such removal from two sites denoted by the letters A and B, as depicted in Figure 1(b) and 1(c). The A-defect is obtained by deleting one carbon atom in the central line of C--C dimers parallel to the armchair border, while the B-defect is created by deletion of a single carbon atom in a secondary line. When the extremities of the strip are attached, the exact location of the defect along $L$ does not affect the structural and electronic properties of the closed untwisted nanoribbons. For the twisted nanoribbons the initial position of the defect site is random, the effects of arbitrary choice hopefully removed after successive geometry optimizations.

\begin{figure}[t]
\centerline{\includegraphics[width=0.47\textwidth]{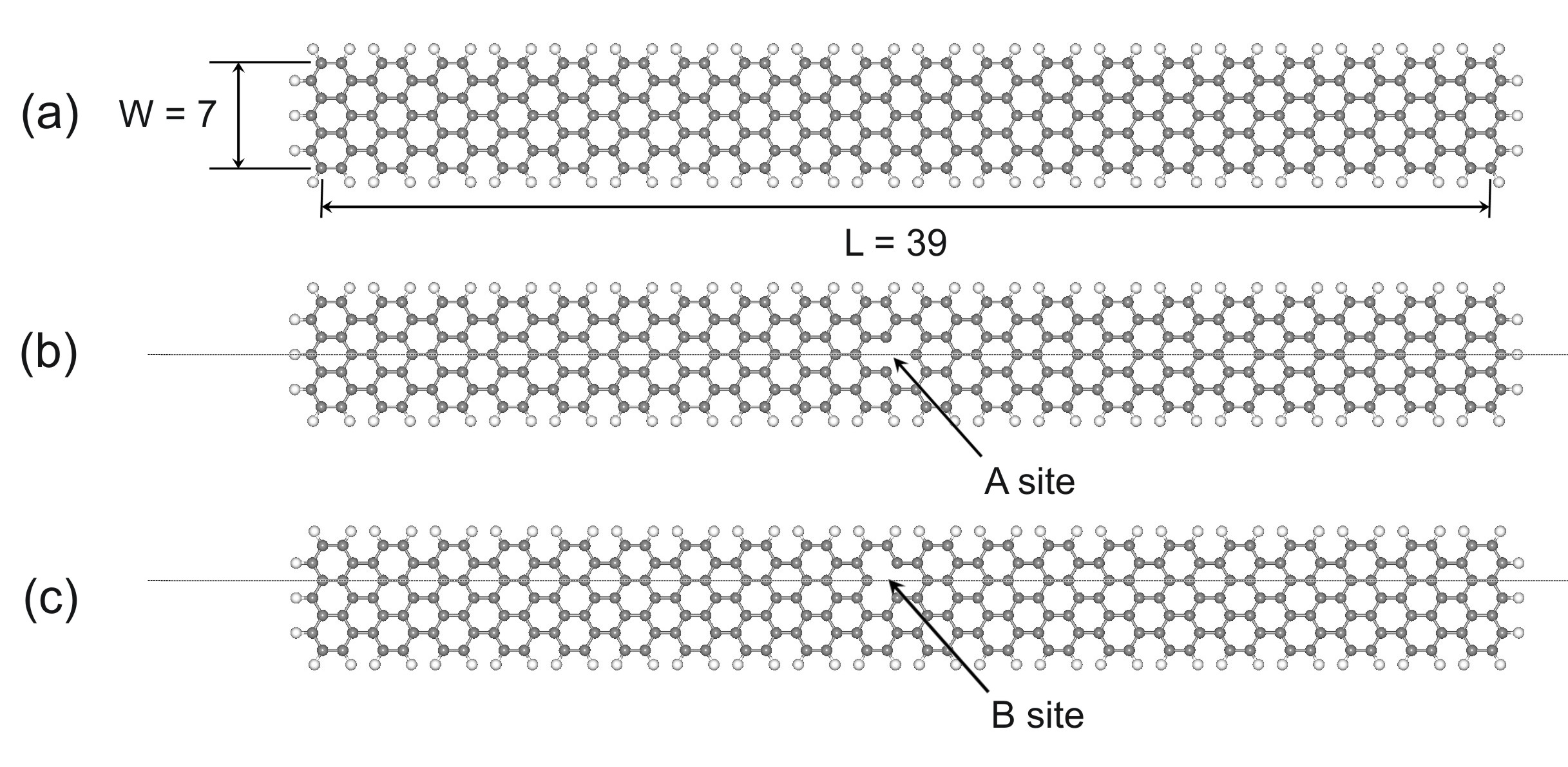}}
\caption{(a) Nanoribbon extracted from a graphene sheet, showing its dimensions $L$ and $W$. (b) The A-defect created by removing a single atom of carbon from the nanoribbon central line (dotted line) of C--C dimers. (b) B-defect created by removing a carbon atom from a secondary line (dotted line).} \label{figure1}
\end{figure}

Fifteen closed nanoribbons were built from the graphene strip,  structures with the number of half-twists $N_t = 0,1,2,3,4$ and the number of defects $N_d = 0,1$, where the $N_d = 1$ case takes into account the two distinct defect sites A and B. Each nanoribbon was used as input to classical molecular mechanics calculations in order to estimate the best structures to be used as starting geometries to compute in the quantum semiempirical framework. For this purpose we have selected the Forcite code and the COMPASS forcefield, which was optimized for condensed-phase atomistic studies, its parameters being derived from \textit{ab initio} calculations. To compute the molecular structures that minimize the classical total energy we proceeded in three stages: first a geometry optimization is made followed by a thermal annealing and new geometry optimization. In the first stage we start from crude structures obtained by simply twisting the graphene rectangle and joining its ends. These structures were optimized obeying some convergence thresholds: total energy variation smaller than $2 \times 10^{-5}$ kcal/mol, maximum force per atom smaller than $10^{-2}$ kcal/mol/nm, and maximum displacement smaller than $10^{-6}$ nm. The optimization algorithm uses a cascade of steepest descent, adjusted basis set Newton-Raphson and quasi-Newton methods. The non-bond interactions included electrostatic and van der Waals terms, with an atom based summation method and cutoff distance of 3.2 nm. The optimized geometries correspond to local minima in the total energy hypersurface of each nanoribbon. The second stage was accomplished in order to achieve geometries corresponding to global minima (or improved local minima). In the thermal annealing calculation, each nanoribbon was submitted to a series of heating and cooling cycles to prevent structural trapping in a suboptimal energy minimum, a very likely outcome in the case of the optimized nanoribbons with a single defect and $N_t>0$. Under such circunstances the defect site position along the length of the nanoribbon can be very important due to the smaller degree of symmetry in comparison with the $N_t=0$ case. Initial and mid-cycle temperatures were chosen to be 300 K and 1000 K, and 100 cycles were computed with 10 heating ramps per cycle and 100 steps per ramp. After each cycle, a geometry optimization using the same criteria for the first stage was carried out. The NVT ensemble was adopted with a time step of 1.0 fs and a Nose thermostat with Q ratio of 1.0. In the third stage, we have carried out for all nanoribbons a new set of computations in order to reinforce our confidence that the obtained geometries are not the results of local mimima and/or used algorithms. We have chosen the same protocol of our previous work \cite{mob11} using reactive molecular dynamics (Brenner-Tersoff type) with different starting velocities and temperatures. This incorporates some stochastic aspects in the tests. The obtained partial structures were then reoptimized to check whether the obtained geometries were consistent with the expected miminum energy configurations.

\begin{figure}[t]
\centerline{\includegraphics[width=0.47\textwidth]{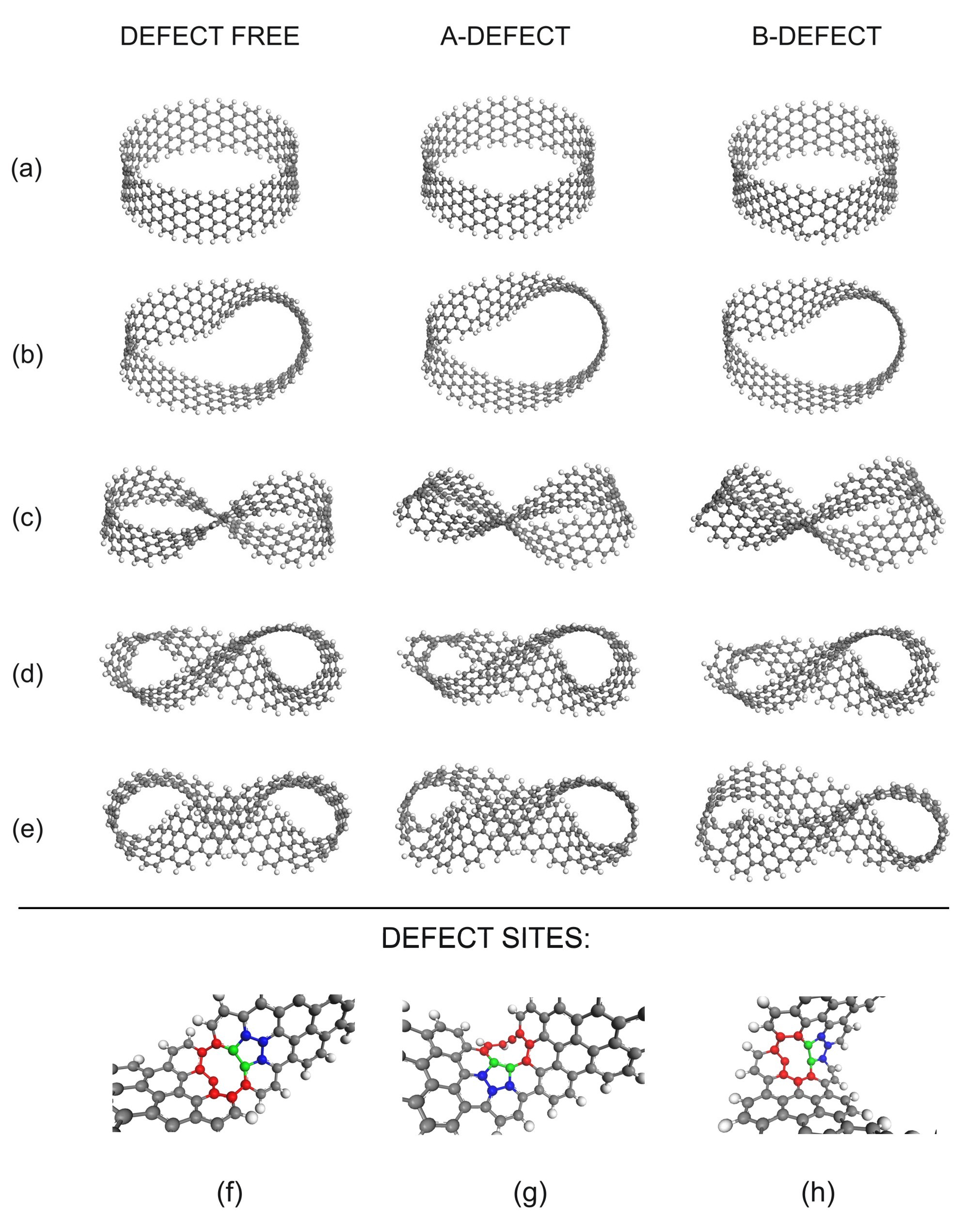}}
\caption{Nanoribbons with geometry optimized at the semiempirical level: (a) untwisted nanoribbon ($N_t=0$), (b) M\"{o}bius nanoribbon ($N_t=1$), (c) $N_t=2$ nanoribbon, (d) $N_t=3$ nanoribbon, and (e) $N_t=4$ nanoribbon. The bottom part of the Figure shows the defect sites for (f) $N_t=1$ A-defect, (g) $N_t=1$ B-defect and (h) $N_t=2$ A-defect nanoribbons. The rings with 9 and 5 carbon atoms at the defect sites are highlighted.} \label{figure2}
\end{figure}

The best geometries computed in the classical mechanics formalism were used to begin a series of semiempirical calculations. The Austin Model 1 (AM1) Hamiltonian implemented in the Gaussian03 code was chosen to perform a new geometry optimization within the restricted Hartree-Fock approximation. There are reports in the literature \cite{mob11a,mob11b,mob11c,mob11d} indicating that both AM1 and first-principles Hartree-Fock calculations show comparable results for cages made of carbon or silicon supporting the applicability of the AM1 semiempirical hamiltonian to investigate all-silicon and all-carbon clusters. With respect to the electronic structure, first principles DFT (Density Functional Theory) calculations tend to overestimate the electronic correlation, underestimating the main energy gap (HOMO-LUMO) of molecular systems in comparison with experiment and semiempirical computations. Indeed, semiempirical CI calculations include some improvements on estimating the electronic correlation energy in comparison with AM1, predicting energy gaps between the ground state and the first excited state intermediary between the DFT and simple AM1 estimates (those tend to follow the trend to predict energy gaps larger than experiment observed in first principles Hartree-Fock methods, which do not take into account electronic correlation effects). Using molecular mechanics, semiempirical and higher-level DFT-B3LYP methods, Xu \textit{et al.} \cite{mob11e} have found the ground-state isomers of large fullerenes (more than 100 atoms). Comparisons between the semiempirical and DFT-B3LYP results show that AM1, PM3, and MNDO semiempirical hamiltonians are notably less accurate for the prediction of the relative energies. On the structural viewpoint, however, bond lengths in carbon nanotubes calculated within semiempirical approaches compare reasonably well with bond lengths found using different DFT exchange-correlation potentials \cite{mob11f}. So we are confident that the structural and electronic structure features presented in this work can be considered, at least, as general trends for the features of real twisted carbon nanoribbons.

Singlet ground states were calculated taking into account the 1760 (1754) electrons of the defect-free (single defect) nanoribbons. Normal modes and the infrared spectra were obtained afterwards and checked for negative frequencies. Finally, the oscillator strengths to estimate the UV-VIS spectra were found using the C.I.(Configuration Interaction) wavefunction approach taking into account the complete active space with contributions from 24 molecular orbitals (HOMO-11 up to LUMO+11). In the end, the C.I. ground state plus the first 127 excited states were acquired and analyzed. The final structures, after the semiempirical geometry optimization of the closed nanoribbons, are presented in Fig.\ 2.

\section{Structural Properties}

A few of the defect-free nanoribbons that minimize the semiempirical total energy exhibit some degree of symmetry. The $N_t=0$ nanoribbon has symmetry point group $D_{20h}$. The M\"{o}bius nanoribbon ($N_t=1$), on the other hand, has symmetry group $C_1$, differing from the corresponding atomic configuration used to start the calculations, which belongs to the $C_2$ point group. The optimized $N_t=2$ nanoribbon has symmetry point group $C_2$ (starting from a $C_2$ configuration), while for the $N_t=3$ (starting configuration with symmetry group $C_3$) and $N_t=4$ (initial symmetry $C_2$) cases the symmetry group is $C_1$. The loss of symmetry in molecular strips is a consequence of the localization of the twists due to the search for a total energy minimum, being previously reported by Mart\'{\i}n-Santamaria and Rzepa \cite{mob3} and in our previous work \cite{mob11}, both using the semiempirical framework. We believe that such twist localization is a general topological feature of all molecular M\"{o}bius systems.

Nanoribbons with A and B defects exhibit structural changes in the carbon-carbon connectivity at the defect sites. The usual hexagonal rings of carbon atoms are replaced there by two adjacent rings with different sizes, the biggest one having nine carbon atoms, while the smallest one has only five carbon atoms, as seen at the bottom part of Fig.\ 2 for the $N_t=1$, A,B-defect (f,g), and the $N_t=2$, A-defect (h) nanoribbons. In A-defect sites, the 9-carbon ring has C--C dimers along both armchair edges of the nanoribbons with $N_t=0,1,4$, but not for the $N_t=2,3$ cases. For the B-defect nanoribbons, the 9-carbon ring has always a C--C--C--C--C pentamer at the edge of the nanoribbon, except for the $N_t=4$ case, where only a C--C dimer stays at the edge. The localization of the 5-carbon ring is variable, being either along the central line of the nanoribbon (A-defect, $N_t=0,1,4$, B-defect, $N_t=2,3$) with no C--C dimer along its borders, or with one C--C dimer along the nanoribbon border (A-defect, $N_t=2,3$, B-defect, $N_t=0,1,4$). The formation of these 9,5-carbon rings involve some degree of pyramidalization at the defect sites and, as we will see later, to a very pronounced increase in the local curvature of the nanoribbons in comparison with their defect-free counterparts.

\vskip1truecm
\begin{table}[h]
\begin{center}
\caption{Heat of formation (in kcal/mol) calculated within the semiempirical formalism for each nanoribbon.} \vskip1truecm
\begin{tabular}{llll}
\hline\hline
\multicolumn{1}{c}{$N_t$} & \multicolumn{1}{c}{Defect-free} & \multicolumn{1}{c}{A-Defect} & \multicolumn{1}{c}{B-Defect} \\
\hline
\multicolumn{1}{c}{0} & \multicolumn{1}{c}{1232.67} & \multicolumn{1}{c}{1405.90} & \multicolumn{1}{c}{1362.23} \\
\multicolumn{1}{c}{1} & \multicolumn{1}{c}{1298.16} & \multicolumn{1}{c}{1458.67} & \multicolumn{1}{c}{1413.09} \\
\multicolumn{1}{c}{2} & \multicolumn{1}{c}{1323.39} & \multicolumn{1}{c}{1545.99} & \multicolumn{1}{c}{1510.54} \\
\multicolumn{1}{c}{3} & \multicolumn{1}{c}{1446.19} & \multicolumn{1}{c}{1594.70} & \multicolumn{1}{c}{1534.67} \\
\multicolumn{1}{c}{4} & \multicolumn{1}{c}{1565.88} & \multicolumn{1}{c}{1705.08} & \multicolumn{1}{c}{1696.53} \\
\hline\hline
\end{tabular}
\end{center}
\end{table}

Under the energetic viewpoint, the heat of formation ($\Delta H_f$) of the twisted nanoribbons increases with the number of half-twists $N_t$, as one can see from Table 1. The defect-free nanoribbons have smaller values of $\Delta H_f$ when compared with the respective values for the A- and B-defect nanoribbons. Starting at 1232.67 kcal/mol ($N_t=0$), $\Delta H_f$ increases up to 1565.88 kcal/mol ($N_t=4$) in an almost linear fashion as a function of $N_t$. For nanoribbons with up to 7 half-twists, it has been shown that the total energy can be approximated as a quadratic function of $N_t$ \cite{mob11}. The A-defect nanoribbons for the same $N_t$ have larger values in comparison with the B-defect ones, being in general larger by about 30-60 kcal/mol. One possible explanation for this difference is that it is more energetically expensive to remove a carbon atom from the central line of C--C dimers in a nanoribbon than from a secondary line. A carbon atom in a secondary line is closer to the edge of the nanoribbon, a place where the energy cost of removing a single atom is lower. However, for $N_t=4$, this difference decreases to less than 10 kcal/mol. It may be possible that for nanoribbons with number of half-twists higher than 4 the difference in the heat of formation between A- and B-defect nanoribbons is very small.

\begin{figure}[t]
\centerline{\includegraphics[width=0.47\textwidth]{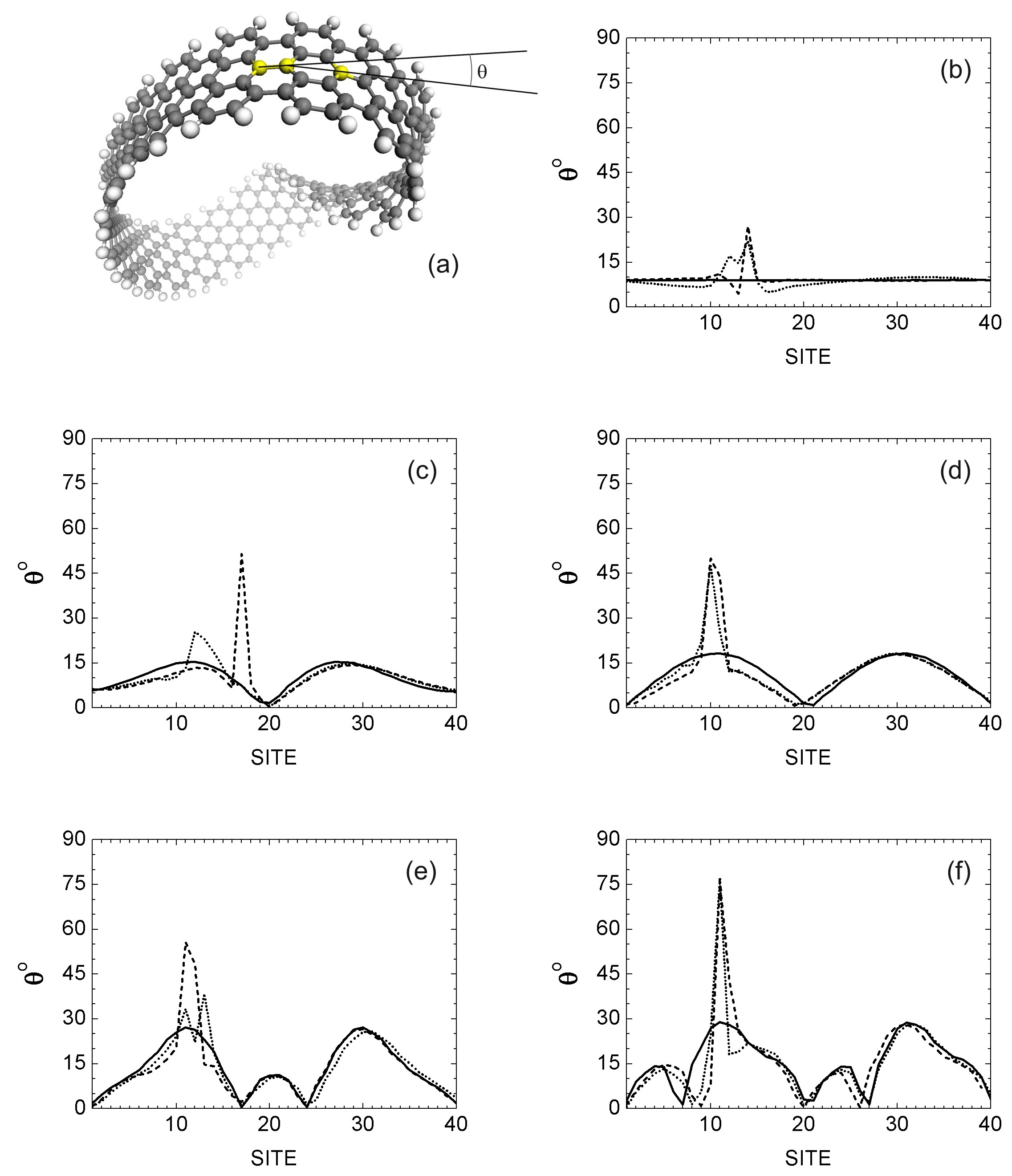}}
\caption{(a) The curvature angle is measured along the central line of atoms of the nanoribbon for three adjacent atoms. Curvature angle along the central line for all nanoribbons: (b) $N_t=0$, (c) $N_t=1$ (M\"{o}bius), (d) $N_t=2$, (e) $N_t=3$, (f) $N_t=4$. Each atom at the vertex of the angle is a site. Solid lines stand for the respective defect-free nanoribbon. Dashed lines stand for the A-defect nanoribbon and dotted lines correspond to the B-defect nanoribbon.} \label{figure3}
\end{figure}

To investigate the effect of the half-twists and defects on the curvature of the closed nanoribbons, we have defined the curvature angle $\theta$, measured along the central line of carbon atoms shown in Fig. 1. This is the angle between the straight lines connecting pairs of contiguous carbon atoms, as shown in Fig. 3(a). For the $N_t=0$ nanoribbon, as one can see from Fig. 3(b), the defect-free configuration has a practically constant value of $\theta$ along its central line of atoms, about 9$^o$ (as expected from a regular polygon with 40 sides). The A-defect nanoribbon shows a noticeable peak of $\theta$, reaching about 27$^o$, followed closely by the B-defect nanoribbon, with 22$^o$. So, the presence of defects tend to create a kink in the nanoribbon. This can be understood in terms of mechanical rigidity: defects lower the mechanical resistance of the nanoribbons to shear stresses created by the process of curving the graphene sheet to form closed strips. This process seems to be reinforced when the closed strips are twisted, as shown in Fig. 3(c,d,e,f).

For the $N_t=1$ nanoribbon (M\"{o}bius case) with no defects, $\theta$ oscillates smoothly between two almost symmetrical regions with pronounced curvature, $\theta_{max} \approx 15^o$, and two regions of low curvature, one with $\theta_{min} \approx 0.17^o$, very flat, and the other with $\theta_{min} \approx 5.8^o$. The insertion of an A-type defect produces a very pronounced increase of $\theta$ in one of the high curvature regions, with $\theta_{max}$ reaching about 51$^o$. The B-defect has not such a strong peak for the curvature angle, reaching about 25$^o$ at most. The second high curvature region remains practically preserved for both the defect-free and the defective nanoribbons. The largest curvature angle in the case of an A-defect for $N_t=1$ occurs because the deletion of a central carbon atom weakens the mechanical resistance of the nanoribbon by creating an empty space (a 9-carbon atoms ring) that crosses its entire width (see Fig. 2(f)). The B-defect, on the other hand, leaves an empty space only in one of the sides of the nanoribbon, with an extra hexagonal ring of carbon atoms reinforcing its structural integrity (see Fig. 2(g)).

This explanation becomes more cogent by looking to what happens with $\theta$ when the nanoribbon has two half-twists. The $N_t=2$ nanostructure presents a 9-carbon ring contiguous to a secondary 5-(6-) carbon ring at its A-(B-) defect site, these rings being parallel to the nanoribbon's width. As the 5-carbon ring has approximately the same mechanical resistance to shear of the 6-carbon one, it is expected that for both A- and B-defect nanoribbons the behaviour of $\theta$ along the central line of carbon atoms should be very similar, as shown in Fig.\ 3(d). The defect-free nanoribbon has two very flat, symmetrical regions with $\theta_{min} \approx 0.89^o$, and two high curvature regions with $\theta_{max} \approx 18^o$. The A- and B-defect nanoribbons have very close peak values of $\theta$, 50$^o$ and 48$^o$ respectively, at one of the high curvature regions observed for the defect-free nanoribbon. The second region of high curvature looks pretty much the same for all structures, as occurred for $N_t=1$ (and occur also for the $N_t=3,4$).

The $N_t=3$ nanoribbon has its $\theta$ profile described in Fig. 3(e). For the nanoribbon with no defect three maxima of $\theta$ can be seen, two of them nearly symmetrical, with $\theta_{max} \approx 27^o$, and a secondary maximum with $\theta_{max} \approx 11^o$. Three minima of $\theta$ exist, two symmetrical flanking the secondary maxima with $\theta_{min} \approx 0.50^o$ and one minimum at the opposite side of the nanoribbon, $\theta_{min} \approx 1.7^o$. The A-defect nanoribbon $\theta$ has a peak reaching $\theta_{max} \approx 56^o$ nearly at the same site of one of the principal maxima of the defect-free structure. For the B-defect, two peaks can be seen in the same region with $\theta_{max} \approx 38^o$. The A-defect site forms a structure with two rings, one with 9 carbon atoms and the other with 6 carbon atoms, across the width of the nanoribbon, while the B-defect site has one 9-carbon ring, two 6-carbon rings and one 5-carbon ring in the same disposition, enhancing its resistance to shear in comparison to the A-defect case, elucidating the highest $\theta_{max}$ observed for the later.

Finally, for $N_t=4$ the mirror symmetry of $\theta$ about the middle site observed in all previous cases ($N_t=1,2,3$) for the free-defect nanoribbons disappears, giving rise to a distinct pattern of alternate maxima and minima (see Fig.\ 3(f)). From there one can see that $\theta$ increases from 2.7$^o$ to 14$^o$, than decreases to 1.4$^o$, rises again to 29$^o$ and tends to return to its initial value of 2.7$^o$, as we move from site 1 to site 20. This sequence repeats itself from sites 21 to 40. The nanoribbons with A- and B-defects show the same general behaviour for $\theta$, reaching practically the same $\theta_{max}$, 76$^o$ and 77$^o$, respectively. These are the highest values for $\theta$ observed for all nanoribbons and one can assume that this is an effect of the high degree of strain in a nanoribbon with such a big number of half-twists and the consequent coiling of the nanoribbon into three dimensions \cite{mob12}.

\begin{figure}[!]
\centerline{\includegraphics[width=0.45\textwidth]{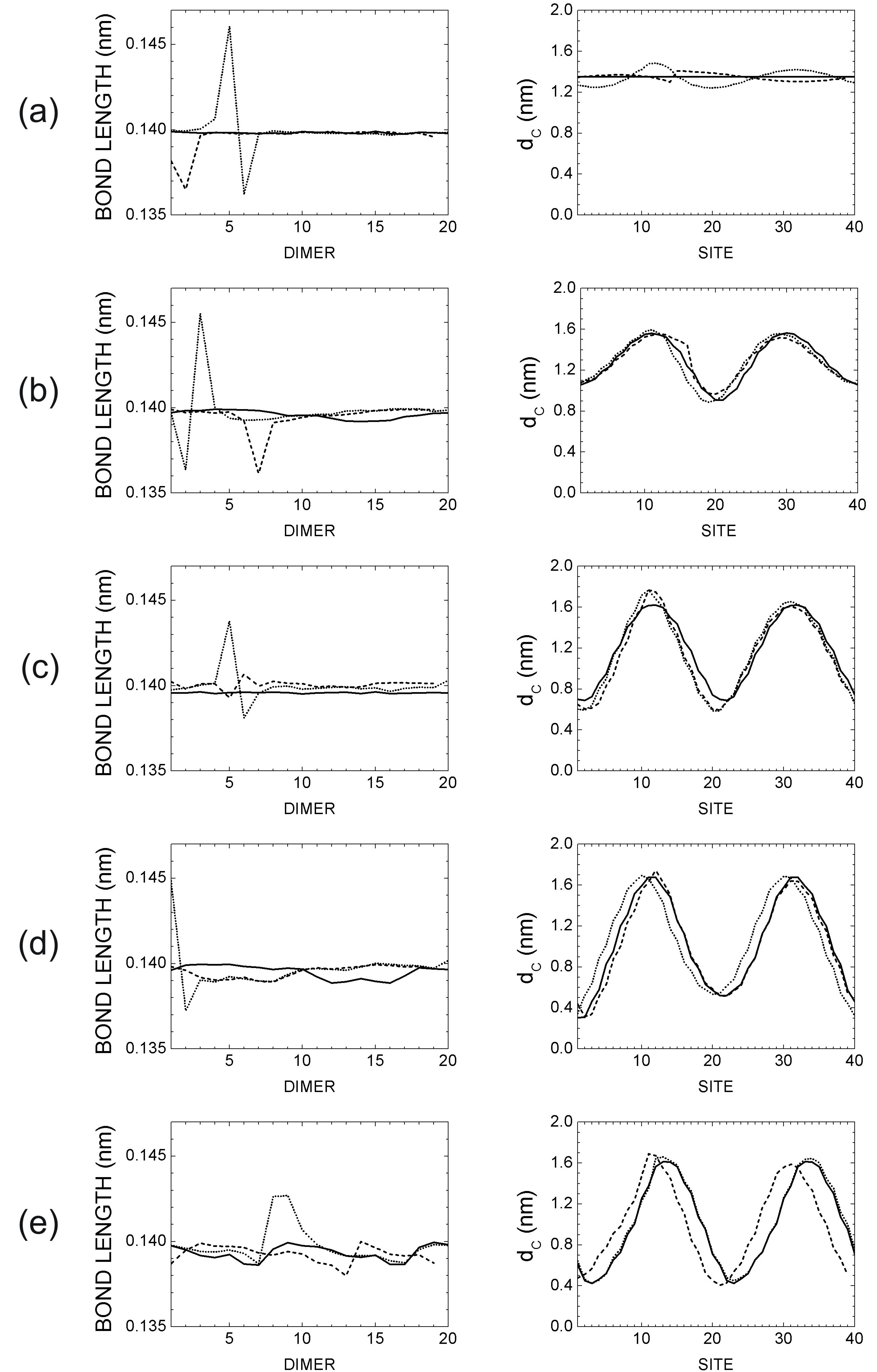}}
\caption{Dimer bond lengths (left) and distances between atom sites located at the central line of the nanoribbon and the nanoribbon centroid (right). The nanoribbons have (a) $N_t=0$, (b) $N_t=1$ (M\"{o}bius), (c) $N_t=2$, (d) $N_t=3$, (e) $N_t=4$. Solid lines, dashed lines and dotted lines indicate the defect-free, A-defect and B-defect nanoribbons, respectively.} \label{figure4}
\end{figure}

Besides the structural analysis using the curvature angle, we have also performed measurements of the bond lengths of all C--C dimers along the central line of atoms for each nanoribbon, as well as the distance between each carbon atom in the central line and the nanoribbon's centroid. The results of these measurements are presented in Fig. 4. One can see that the bond lengths for A-defect nanoribbons oscillate more strongly in a certain region for each $N_t$, which corresponds to the respective defect site. More pronounced peaks in bond length can be seen for $N_t=0,1,2,3$, while for $N_t=4$ there is a smaller, flat peak. The bond length along the nanoribbon ranges from about 0.136 nm to 0.146 nm close to the defect site and is kept practically constant -- about 0.140 nm -- elsewhere. So the presence of defects produce a C--C bond length variation of about 4.3\% at most in the central line of atoms. For the B-defect nanoribbons, the bond length variations are smaller due to the localization of the defect in a secondary line, parallel to the central line of C--C dimers. The bond length for B-defect nanoribbons tends to decrease near the defect, and its variation is smaller for nanoribbons with $N_t=2,3,4$. In the right side of Fig. 4, the distance to the nanoribbon's centroid, $d_c$, is plotted for all structures. For the $N_t=0$ case, $d_c$ is constant in the defect-free nanoribbon and oscillates slightly in the B-defect one. The A-defect nanoribbon $d_c$ oscillates more remarkably near the defect site. For $N_t=1,2,3,4$, $d_c$ oscillates with maxima in the regions of highest curvature angle (compare with Fig. 3). The amplitude of oscillation is larger for greater values of $N_t$. The defects does not seem to affect remarkably the behavior of $d_c$, except for a small shift in the localization of its maxima, as can be seen clearly for $N_t=3,4$. Finally, the length of the central line of atoms $L_{CL}$ decreases from 8.49 nm to 8.38 nm for the defect-free nanoribbon as $N_t$ increases from 0 to 4. For the A-defect nanoribbon, $L_{CL}$ starts at 8.48 nm ($N_t=0$), increases to 8.53 nm ($N_t=1$), decreases to 8.38 nm ($N_t=2$), 8.33 nm ($N_t=3$) and increases again to 8.49 nm ($N_t=4$). In the B-defect nanoribbon, $L_{CL}$ decreases from 8.45 nm ($N_t=0$) down to 8.38 nm ($N_t=3$) and then increases to 8.44 nm ($N_t=4$).

\section{Electronic States, UV-VIS and Infrared Spectra}

Subsequent to the geometry optimization at the semiempirical level, the electronic eigenstates were calculated for each nanoribbon. In Fig. 5 one can see the HOMO (Highest Occupied Molecular Orbital) and LUMO (Lowest Unoccupied Molecular Orbital) isosurfaces corresponding to a wavefunction amplitude of $\pm 0.02$ for $N_t=0,1$. The $N_t=2,3,4$ HOMO and LUMO orbitals were also obtained, but for the sake of brevity we have focused only in the $N_t=0,1$ cases as they are representative of the general features observed for these orbitals in all configurations investigated in this work.

The HOMO states present phase alternation along the width of each nanoribbon for the $N_t=0,1$ defect-free and A-defect structures. For the B-defect nanoribbon, the pattern of alternating phases of the HOMO orbital seems to be rotated with respect to its armchair edge. LUMO orbitals, on the other hand, exhibit phase alternance along the length of all nanoribbons, except for the B-defect nanoribbon, which behaves in a similar fashion to the observed for the HOMO orbital. Both patterns of phase alternance associated to the defect-free geometries match the observations we have made in a previous paper \cite{mob11} considering $L=29$, $W=5$ twisted nanoribbons. For $N_t=0$, defect-free structures, the molecular orbitals are delocalized. The twisting process, on the other hand, tends to localize the molecular orbitals in a region of the molecular strip. When A-,B-defects are inserted, distinct effects on the HOMO (LUMO) localization appear. In the case of an A-defect nanoribbon with $N_t=0,1$, both HOMO and LUMO states confine electrons outside the region of the defect. On the other hand, for the B-defect case, electrons in the frontier orbitals show some degree of localization close to the defect sites. As the B-defect is localized near the edge of the nanoribbon, we presume that the distinct behaviour of the HOMO and LUMO orbitals is due to the stronger interaction -- in comparison to the A-defect structure --  between dangling bond states at the defect site with electronic states along the edge of the nanoribbon.

\begin{figure}[!]
\centerline{\includegraphics[width=0.45\textwidth]{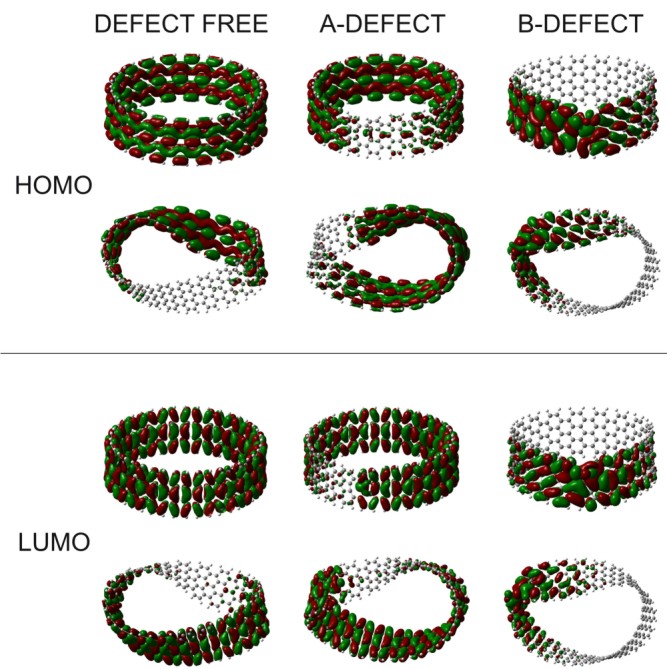}}
\caption{HOMO (top) and LUMO (bottom) orbitals for the $N_t=0,1$ nanoribbons. The first and third (second and fourth) rows correspond to the $N_t=0$ ($N_t=1$) case. The first column refers to the defect-free nanoribbons, while the second and third columns show the molecular orbitals for the A- and B-defect nanoribbons, respectively.} \label{figure5}
\end{figure}

Eigenenergies calculated within the semiempirical method for the HOMO and LUMO electronic states are presented in Fig. 6, top-left. The defect-free nanoribbons show small variation of these energies as $N_t$ is switched from 0 to 4, with an average value of about -7.45 eV for the HOMO and -2.03 eV for the LUMO. The LUMO energy decreases almost linearly when $N_t$ increases from 0 to 4, while the HOMO energy oscillates somewhat, increasing from $N_t=0$ to $N_t=1$, decreasing from $N_t=1$ to $N_t=2$, increasing from $N_t=2$ to $N_t=3$ and decreasing again from $N_t=3$ to $N_t=4$. The A-defect nanoribbons have a more pronounced variation of the HOMO and LUMO energies as the number of half-twists changes. For the HOMO states, the energy starts at -7.46 eV ($N_t=0$), decreases slightly to -7.48 eV ($N_t=1$) and increases by almost 0.16 eV, reaching about -7.32 eV ($N_t=2$). Than it decreases to -7.44 eV ($N_t=3$) and again to -7.49 eV ($N_t=4$). The LUMO energy behaves in a reverse way with respect to $N_t$ in comparison with the HOMO energy values. It decreases when the HOMO increases and increases otherwise: -2.01 eV ($N_t=0$), -2.04 eV ($N_t=1$), -2.30 eV ($N_t=2$), -2.16 eV ($N_t=3$), and -2.07 eV ($N_t=4$). Finally, for the B-defect structure we have some oscillation of the HOMO, LUMO eigenenergies as the value of $N_t$ is altered. These oscillations seem to mimic the oscillations observed for the A-defect nanoribbon in reverse. So when the HOMO energy for the A-defect nanoribbon is augmented by adding a twist, the corresponding value for the B-defect nanoribbon tends to get diminished. For the HOMO energies the values are -7.30 eV, -7.27 eV, -7.44 eV, -7.39 eV, and -7.48 eV, for $N_t=0,1,2,3,4$, respectively. LUMO energies are -2.24 eV, -2.25 eV, -2.12 eV, -2.15 eV, and -2.17 eV for $N_t=0,1,2,3,4$ in this order. Therefore for all cases, except for $N_t=2$, the HOMO (LUMO) energies for the B-defect nanoribbon are larger (smaller) than the corresponding values for the A-defect nanoribbon.

The HOMO-LUMO gap $E_g$ is shown in Fig.\ 6, top-right. As the number of half-twists $N_t$ changes for the defect-free nanoribbon, $E_g$ exhibits a very small variation, with minimum value of 5.41 eV for $N_t=1,34$ and maximum value of 5.43 eV ($N_t=0,2$), a difference of only 20 meV. This could be compared with the data obtained for $L=29$, $W=5$ nanoribbons \cite{mob11}, with HOMO-LUMO gaps of 4.65 eV, 4.58 eV, 4.66 eV, 4.56 eV, 4.55 eV for $N_t=0,1,2,3,4$, respectively, the difference between the minimum and maximum values being 110 meV. It seems that by increasing the size of the nanoribbon, the HOMO-LUMO gap tends to increase and becomes somewhat unsensitive to the variations in the number of half-twists.

When a single defect is taken into account, a very pronounced variation is observed for $E_g(N_t)$ as $N_t$ is switched from 0 to 4. For the A-defect nanoribbons, $E_g$ starts at 5.45 eV ($N_t=0$), than decreases a little bit to 5.44 eV ($N_t=1$). For $N_t=2$ a sharp decrease occurs, $E_g$ becoming 5.02 eV. For $N_t=3,4$, the HOMO-LUMO gap energy increases to 5.28 eV and 5.42 eV, respectively. The $E_g$ variation is 430 meV between the minimum ($N_t=2$) to the maximum ($N_t=0$). The B-defect nanoribbon also shows a large variation of $E_g(N_t)$ in comparison with the defect-free case: 300 meV between $E_g(N_t=2)=5.32$ eV and $E_g(N_t=1)=5.02$ eV. Other values of $E_g$ are 5.06 eV ($N_t=1$), 5.24 eV ($N_t=3$), and 5.31 eV ($N_t=4$). One can note that the minimum value of $E_g$ for the A-defect nanoribbon occurs for the same number of half-twists $N_t=2$ the B-defect nanoribbon $E_g$ exhibits its highest value.

To investigate the excited states we have performed a full configuration interaction (CI) computation for each nanostructure. In order to build the CI determinants, the CI active space was formed using 24 molecular orbitals from HOMO-11 to LUMO+11. The singlet ground state is obtained by filling the 600 (598) lowest energy eigenstates with two electrons per level, leading to a total of 1200 (1196) electrons for the defect-free (A-,B-defect) nanoribbons. A total of 127 excited states were then calculated and the oscillator strengths for optical transitions between the singlet ground state and these excited states were computed to obtain UV/Visible absorption spectra. The bottom of Fig.\ 6 shows the CI gaps obtained by taking the energy difference between the first excited state T1 and the ground state GS (lines with open symbols) and by taking the energy difference between the first excited state with transition oscillator strength different from zero A1 and the ground state (first allowed optical transition represented by the lines with solid symbols). In all cases, the first excited state was a triplet, with optical transition from the singlet ground state forbidden by the spin selection rule. The GS $\rightarrow$ T1 gaps for the defect-free nanoribbons (solid line, open squares) vary from 2.49 eV ($N_t=1$, minimal) to 2.63 eV ($N_t=0$, maximum), a difference of 140 meV. The $N_t=2,3$ nanoribbons have GS $\rightarrow$ T1 transition energies very close to the $N_t=0$ case, 2.60 eV and 2.61 eV, respectively. The $N_t=4$ GS $\rightarrow$ T1 transition energy is 2.52 eV, closer to the $N_t=1$ structure. On the other hand, the optically allowed transitions GS $\rightarrow$ A1 (solid line, solid squares) have higher values, but follow the same qualitative trend of the GS $\rightarrow$ T1 energies, decreasing from 3.00 eV ($N_t=0$, corresponding to violet color with wavelength 410 nm) to 2.67 eV ($N_t=1$, corresponding to blue color with wavelength 460 nm), increasing to 2.76 eV ($N_t=2$), increasing again to 2.80 eV ($N_t=3$) and decreasing to 2.69 eV ($N_t=4$). The overall variation in energy is 330 meV between the $N_t=1$ and $N_t=0$ nanoribbons.

\begin{figure}[!]
\centerline{\includegraphics[width=0.47\textwidth]{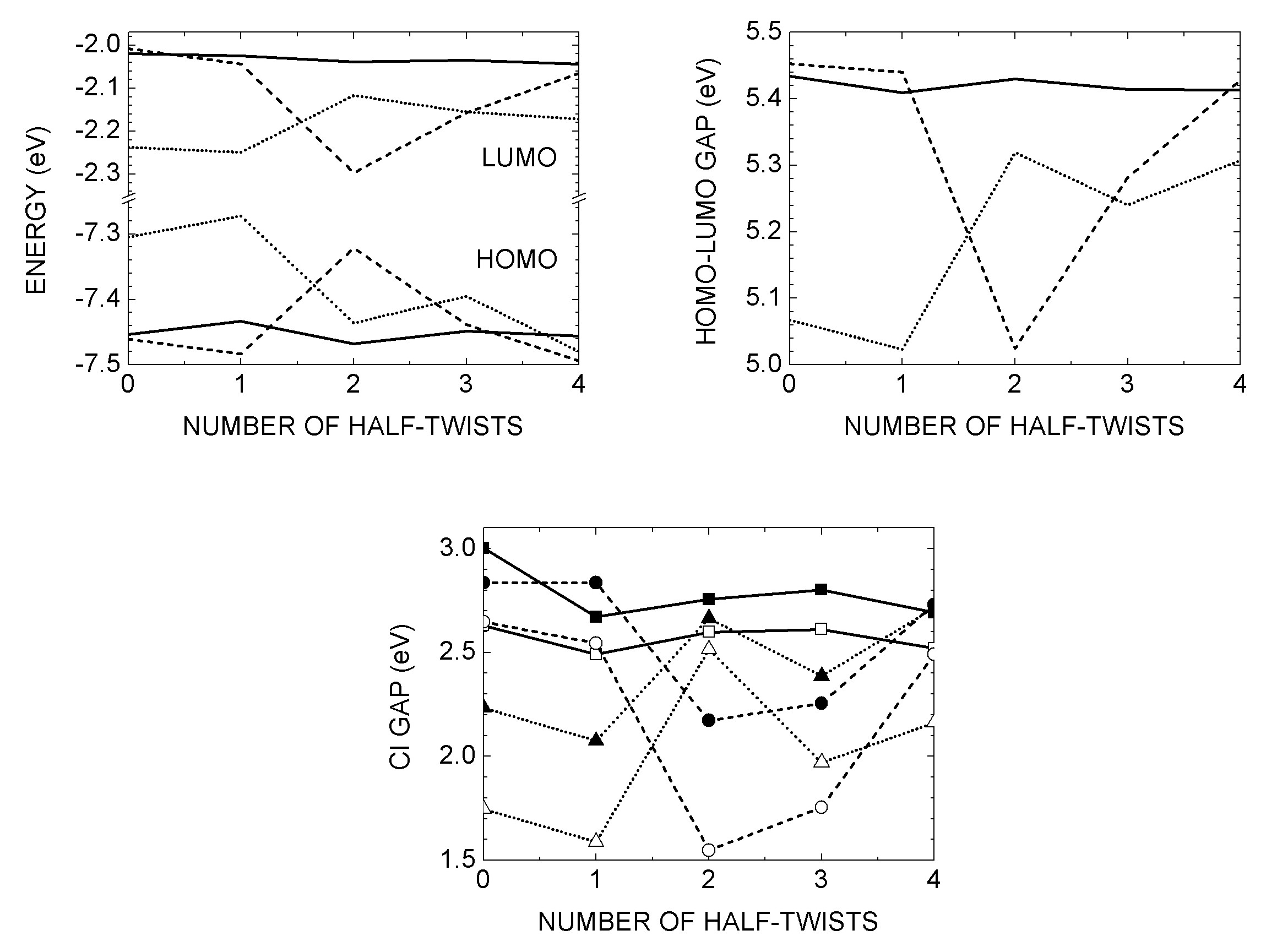}}
\caption{Top, left: HOMO and LUMO energy levels for the $N_t=0,1,2,3,4$ nanoribbons. Top, right: HOMO-LUMO energy gaps. For both plots, the solid lines indicate the defect-free case, while the dashed lines and dotted lines refer to the A- and B-defect nanoribbons, respectively. Bottom: energy gaps calculated within the semiempirical Full CI approach, with the curves with open squares, circles and triangles corresponding to the difference of energy between the ground state (singlet) and the first excited state (in all cases, triplet). Curves with solid squares, circles and triangles correspond to the first allowed optical transition from the singlet ground state to a singlet excited state. Solid, dashed and dotted lines depict, in respective order, the defect-free and the A-, B-defect nanoribbons.} \label{figure6}
\end{figure}

The deletion of a single carbon atom to form an A-defect or a B-defect enhances the variation of the GS $\rightarrow$ T1 and GS $\rightarrow$ A1 CI gaps. Looking to the GS $\rightarrow$ T1 transition for the A-defect nanoribbon (dashed lines, open squares), one can see the GS $\rightarrow$ T1 energy beginning at 2.65 eV ($N_t=0$), decreasing to 2.55 eV ($N_t=1$), decreasing again (and sharply) to 1.55 eV ($N_t=3$), and then increasing to 1.75 eV ($N_t=4$) and 2.49 eV ($N_t=4$). For the optically active transitions GS $\rightarrow$ A1 the energy gaps are 2.84 eV ($N_t=0,1$, violet color with wavelength of 440 nm), 2.17 eV ($N_t=2$, yellow color with wavelength of 570 nm), 2.26 eV ($N_t=3$, green color, wavelength 550 nm), and 2.73 eV ($N_t=4$, blue color, wavelength 450 nm). When a B-defect is formed, the CI gaps as functions of $N_t$ show a behavior contrary to the observed for the A-defect case: when the CI energy of the first increases as one half-twist is added, the corresponding CI energy for the later decreases, and vice-versa, with the exception when one switches from $N_t=3$ to $N_t=4$. The B-defect nanoribbon GS $\rightarrow$ T1 energy gaps are 1.75 eV, 1.59 eV, 2.52 eV, 1.97 eV, and 2.16 eV, for $N_t=0,1,2,3,4$ in respective order. For the GS $\rightarrow$ A1 optically allowed transitions, we find 2.23 eV ($N_t=0$, yellow color, wavelength 560 nm), 2.08 eV ($N_t=1$, orange color, wavelength 600 nm), 2.67 eV ($N_t=2$, blue color, wavelength 460 nm), 2.39 eV ($N_t=3$, green color, wavelength 520 nm), and 2.71 eV ($N_t=4$, blue color, wavelength 460 nm). So it seems to be possible to control the color in the visible spectrum a nanoribbon absorbs by changing the number of half-twists \cite{mob11}. The addition of defects, on the other side, seems to increase the range of wavelengths in the visible spectrum accessible to optical absorption in comparison with the defect-free geometry.

After obtaining the oscillator strengths for all transitions  in our CI calculations, we have plotted the UV/Visible absorption spectra for all nanoribbons, as shown in Fig. 7. A set of Lorentzian curves with amplitudes proportional to the oscillator strengths for each GS $\rightarrow$ excited state (ES) transition and fixed widths of 0.038 eV was used in the plots. The absorption curves for the defect-free nanoribbons are shown in Fig. 7(a),(b),(c),(d),(e) for $N_t$ varying from 0 to 4. When $N_t=0$ (Fig. 7(a)), three remarkable peaks appear at 2.98 eV (violet color, wavelength of 420 nm), 3.73 eV and 4.11 eV (the first energy within the near ultraviolet range and the second within the ultraviolet B range). A very small peak also appears for an energy of 4.86 eV (corresponding to a photon in the ultraviolet C range). Nanoribbons with one or more half-twists show a richer absorption spectrum, with more peaks in comparison with the $N_t=0$ case. For $N_t=1$ (Fig. 7(b)), the first significant peak appears at 2.88 eV, followed by peaks at 3.68 eV, 3.83 eV and 4.10 eV. When $N_t=2$ (Fig. 7(c)), the main absorption peaks appear at 2.94 eV, 3.88 eV (with a series of four smaller peaks between the two first ones), and 4.06 eV. In the $N_t=3$ nanoribbon (Fig. 7(d)), one can see that the single absorption peak of smallest energy observed in the $N_t=0,1,2$ cases is replaced by two peaks, one small at 2.81 eV and the other one, bigger than the first at 2.98 eV. A single small peak appears at 3.39 eV followed by a series of close absorption peaks between 3.84 eV and 4.17 eV. At last, for $N_t=4$ (Fig. 7(e)), a single absorption peak appears at 2.86 eV and a structure of close peaks with almost the same height is revealed in the 3.84 eV -- 4.17 eV energy range. Smaller absorption peaks also can be noted at 4.53 eV and 4.65 eV.

\begin{figure}[!]
\centerline{\includegraphics[width=0.47\textwidth]{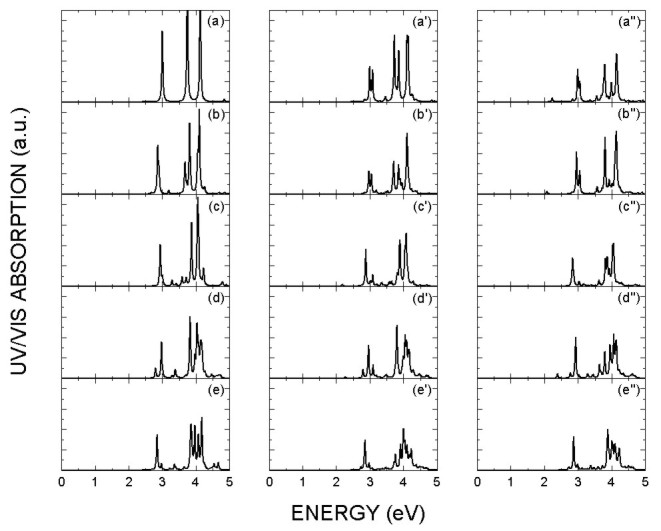}}
\caption{UV/Visible spectra of the nanoribbons estimated by using the oscillator strengths for electronic transitions between the singlet ground state and 127 excited states. (a),(b),(c),(d),(e): defect-free nanoribbons with $N_t=0,1,2,3,4$, respectively. (a'),(b'),(c'),(d'),(e'): $N_t=0,1,2,3,4$, respectively, for the A-defect nanoribbons. (a''),(b''),(c''),(d''),(e''): $N_t=0,1,2,3,4$ for B-defect nanoribbons.} \label{figure7}
\end{figure}

The A-defect nanoribbons have their absorption spectra shown in Fig. 7 (a'), (b'), (c'), (d'), and (e'). For the $N_t=0$ configuration (Fig. 7(a')), five peaks are clearly visible, two near 3 eV, two near 3.8 eV, and one near 4.1 eV. The last peak, after a closer look, exposes itself as a pair of nearby peaks. Therefore it seems that the addition of the defect splits the three sharp peaks at 2.98 eV, 3.73 eV, and 4.11 eV observed in the $N_t=0$ defect-free nanoribbon. Looking now to the same nanoribbon $N_t=0$ with a single B-defect (Fig. 7(a'')), the splitting of absorption peaks is apparent only for the peaks around 3 eV, and a very small (but yet noticeable) absorption peak occurs at 2.23 eV, an energy equivalent to a green light photon with wavelength of about 560 nm. A similar peak also appears at 2.10 eV ((Fig. 7(b'')), $N_t=1$, yellow-orange color) and 2.38 eV ((Fig. 7(d'')), $N_t=3$, green color), but it is absent in the $N_t=2,4$ B-defect nanoribbons ((Fig. 7(c''),(e'')). In the A-defect nanoribbon with $N_t=2$ (Fig. 7(c')), a very small absorption peak occurs for an energy of 2.18 eV (yellow color) and in the $N_t=3$ case (Fig. 7(d')), we have another small peak at 2.25 eV. A common feature of all nanoribbons with defects is a larger number of peaks in comparison with their defect-free counterparts, but with smaller absorption intensities.

\begin{figure}[!]
\centerline{\includegraphics[width=0.45\textwidth]{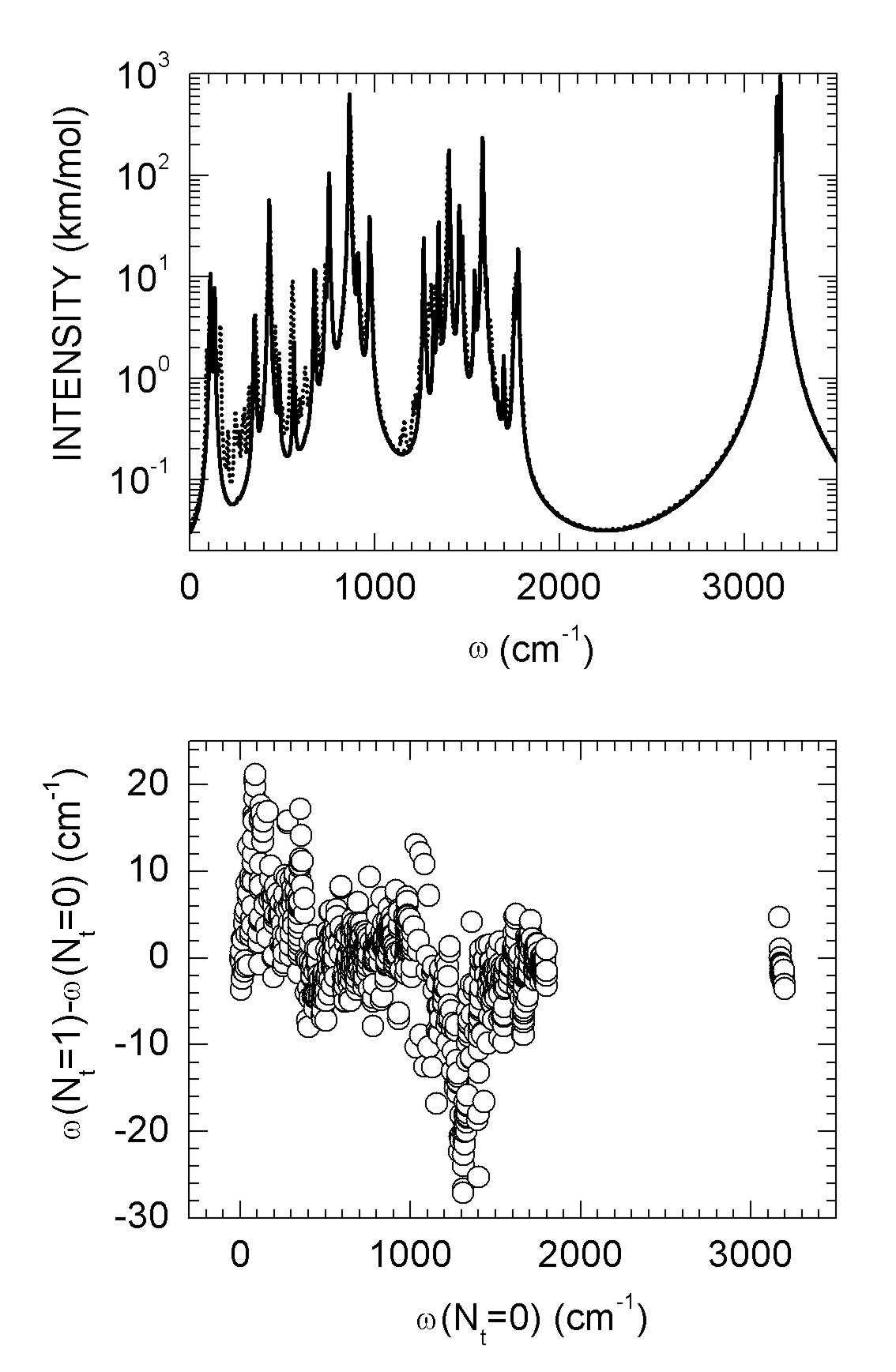}}
\caption{Top: IR spectra for the $N_t=0$ (solid line) and $N_t=1$ (dotted line) defect-free nanoribbons. Bottom: difference of frequencies between corresponding normal modes of $N_t=1$ and $N_t=0$ defect-free nanoribbons (circles). The zero-frequency was shifted for better visualization} \label{figure8}
\end{figure}

Next to the calculation of the electronic absorption spectra in the UV/Vis\-i\-ble range, we now comment the infrared spectra calculated for the defect-free nanoribbons with $N_t=0,1$ (Fig. 8). We do not show the infrared spectra for the other nanoribbons (defect-free with more half-twists, A-, B-defect) because they resemble very closely the spectra we are about to discuss. Indeed, as shown in the top of Fig. 8, the infrared spectra for the $N_t=0$ and $N_t=1$ nanoribbons are so similar that their differences can be seen only when we plot it using a logarithmic scale. The $N_t=1$ IR peaks (dotted line) appear practically at the same energies of the $N_t=0$ case (solid line), with some low intensity peaks exclusive of the $N_t=1$ nanoribbon observable only for the 200-400 cm$^{-1}$, 590-640 cm$^{-1}$, and 1100-1200 cm$^{-1}$ frequency ranges. A comparison of the normal mode frequencies $\omega$ for each nanoribbon was carried out by plotting the difference between the corresponding normal mode frequencies $\omega(N_t=1)-\omega(N_t=0)$ versus $\omega(N_t=0)$, as shown in the bottom part of Fig. 8. A corresponding normal mode is defined as follows: first we enumerate in crescent order of frequency the normal modes of each nanoribbon; second, as both nanoribbons have the same number of vibrational eigenstates (1080 total), we can make an one-to-one correspondence between them. For frequencies between 0 and 1000 cm$^{-1}$, the $N_t=1$ nanoribbon has normal modes with frequencies a little bit higher than the frequencies for the untwisted geometry, with maximum difference of about 21 cm$^{-1}$. In the frequency range 1000-1600 cm$^{-1}$ this trend is reversed, with the $N_t=1$ frequencies being in general lower than the $N_t=0$ ones, with minimum difference of about -27 cm$^{-1}$. Between 1600 and 1800 cm$^{-1}$ and for the highest frequency modes around 3100-3200 cm$^{-1}$, the corresponding normal modes of both nanoribbons, $N_t=0,1$, have very close frequencies. The mean value of the difference $\omega(N_t=1)-\omega(N_t=0)$ was -0.60 cm$^{-1}$, with 6 cm$^{-1}$ of standard deviation. So the normal modes of the $N_t=1$ nanoribbon tend to have slightly smaller frequencies than the corresponding normal modes of the $N_t=0$ structure.

\section{Conclusions}

Recent advances in the synthesis of molecules with half-twists and M\"{o}bius topology stimulate the investigation of their structural, electronic and optical properties. At the same time, the astounding properties of graphene are driving huge research efforts worldwide. One of the possible nanostructures derived from graphene is precisely a closed-loop twisted nanoribbon. In this paper we pointed out some results of classical and quantum semiempirical simulations of graphene-derived twisted nanoribbons with length $L=39$ and width $W=7$ (parameters defined as in Ref.\ \cite{mob11}) with defects created by the deletion of a single carbon atom in two distinct sites, comparing their characteristics with the corresponding defect-free nanoribbons. Hydrogen atoms were added to passivate dangling bonds (except at the defect sites).

After obtaining the best geometries for the nanoribbons using a combination of classical dynamics and classical annealing, a set of closed nanoribbons with number of half-twists ($N_t$) varying from 0 to 4, defect-free and with two types of defect (A and B) was submitted to a process of geometry optimization using the quantum semiempirical Hamiltonian AM1. In defective nanoribbons, the pattern of hexagonal carbon rings is replaced at the defect location by two adjacent rings with 9 and 5 carbon atoms (9-5 rings). The spatial disposition of the 9-5 rings in the nanoribbons seems to depend on the number of half-twists as well as on the type of defect and contribute, through the pyramidalization of C-C bonds, to increase local curvature. Structural features of the carbon nanoribbons were evaluated and it was demonstrated that the insertion of defective sites significantly increases the local curvature angle $\theta$. In general, A-defect nanoribbons have more local curvature at the defect site in comparison with the B-defect ones. C-C bond lengths, however, do not change meaningfully when one compares the defective and defect-free structures among themselves (less than 4\% of difference).

Frontier molecular orbitals are delocalized for $N_t=0$, defect-free nanoribbons, and localized in the twisted (specially in the defective) nanoribbons. By inserting A- (B-) defects, the HOMO and LUMO states have larger amplitudes in regions apart from (at) the defect-site. The HOMO-LUMO gap of the A-defect nanoribbons increases (decreases) when the HOMO-LUMO gap of the B-defect nanoribbons decreases (increases) as functions of $N_t$. The same happens for the CI gaps. In the defect-free nanostructures, the first optically active transitions (with non-zero oscillator strength) involve photon energies corresponding to violet ($N_t=0$) and blue ($N_t=1,2,3,4$) colors. Defective nanoribbons, on the other hand, have the first optically active transitions with energies ranging from 2.08 eV (B-defect, $N_t=1$, orange color) to 2.84 eV (A-defect, $N_t=0,1$, violet color). The most intense UV/Visible absorption peaks were observed in the ultraviolet range, with smaller peaks in the visible spectrum. Contrasting the defect-free nanoribbons absorption lines with the A-, B-defect ones reveals a richer structure of peaks for the later. Adding half-twists also contributes to increase the number of peaks, in some cases through splitting (for example, the $N_t=3$ defect-free nanoribbon). Distinct nanoribbons have very similar infrared spectra, even in the presence of A or B defects.

\textbf{Acknowledgements}
EWSC, VNF, ELA, and DSG are senior CNPq researchers. The first two and the later received financial support from CNPq--Rede NanoBioestruturas, project n. 555183/2005-0. EWSC received financial support from CNPq--process numbers 478885/2006-7, 482051/2007-8, and 304338/2007-9. DGS and FS also acknowledge support from FAPESP. SGS was sponsored by a graduate fellowship from CNPq and CAPES.

\end{document}